\documentclass[12pt]{article}
\usepackage{amssymb,amsmath,graphicx}
\newcommand{\be}{\begin{equation}}
\newcommand{\ee}{\end{equation}}
\newcommand{\bea}{\begin{eqnarray}}
\newcommand{\eea}{\end{eqnarray}}
\newcommand{\beas}{\begin{eqnarray*}}
\newcommand{\eeas}{\end{eqnarray*}}

\def \PRL {{Phys. Rev. Lett.}}

\def \PR {{Phys. Rev.}}

\begin{document}

\begin{titlepage}
\pagestyle{empty}
\begin{center}
{\Large\bfseries
Thermal diffractive corrections to Casimir energies}\\
\vskip .1in
{\Large\bfseries ~}\\

\vspace{.2in}
{\large DANIEL KABAT\footnote{daniel.kabat@lehman.cuny.edu}} and
{\large DIMITRA KARABALI\footnote{dimitra.karabali@lehman.cuny.edu}} \\
\vskip .2in
{\itshape Department of Physics and Astronomy\\
Lehman College of the CUNY\\
Bronx, NY 10468}\\

\fontfamily{cmr}\fontsize{12pt}{16pt}\selectfont
\vspace{.8in}
{\large\bf Abstract}
\end{center}
We study the interplay of thermal and diffractive effects in Casimir
energies.  We consider plates with edges, oriented either parallel or
perpendicular to each other, as well as a single plate with a slit.
We compute the Casimir energy at finite temperature using a formalism in which the diffractive effects are encoded in a lower dimensional 
non-local field theory that lives in the gap between the plates.  The
formalism allows for a clean separation between direct or geometric
effects and diffractive effects, and makes an analytic derivation of the temperature
dependence of the free energy possible.  At low temperatures, with Dirichlet
boundary conditions on the plates, we find that diffractive effects make a correction to the
free energy which scales as $T^6$ for perpendicular plates, as $T^4$ for slits,
and as $T^4 \log T$ for parallel plates.

\end{titlepage}

\pagestyle{plain}
\setcounter{page}{2}
\setcounter{footnote}{0}
\renewcommand\thefootnote{\mbox{\arabic{footnote}}}

\section{Introduction\label{sect:intro}}

The Casimir effect is famous as a prototype for the influence of boundary conditions
in quantum field theory.  The original Casimir effect described the interaction
between two infinite parallel conducting plates due to vacuum fluctuations of the
electromagnetic field \cite{Casimir}.  Since that pioneering work many variants
of the effect have been studied.  For recent reviews see \cite{reviews}.

It is interesting to ask how the Casimir energy is modified when the plates have boundaries, either apertures or edges.  That is, it is interesting
to ask how diffractive effects correct the Casimir energy.  We studied this in
\cite{Kabat:2010nm} using a formalism which we will review below.
An advantage of our formalism is that it allows for a clean separation between direct or geometrical effects associated with the plates,
and diffractive effects associated with the plate boundaries. In \cite{Kabat:2010nm} we considered several geometries: two perpendicular plates separated by a gap,
a single plate with a slit in it, and two parallel plates, one of which is semi-infinite. For other 
approaches to analyzing the Casimir energy in such geometries see \cite{MIT,Milton2,Gies}.

In the present paper we extend our results to finite temperature.  One of our motivations is to obtain an analytic understanding of the non-trivial correlation between geometry and temperature
found in \cite{KlingGies,GiesWeber} using worldline Monte Carlo techniques. Although the high temperature limit of the Casimir energy obeys a well understood, linear dependence on temperature, the low temperature limit is much more subtle and depends crucially on the global configuration of the plates.

By way of outline, in section \ref{sect:effective} we set up the formalism at finite temperature
and collect some useful preliminary results.  We study the behavior at low temperature in section
\ref{sect:low}, with perpendicular plates in section \ref{sect:perp}, slits in section \ref{sect:slit},
and parallel plates in section \ref{sect:parallel}.  We conclude in section \ref{sect:conclusions}.  Appendix
\ref{sect:thermo} collects some useful results on the partition function of an ideal gas.

\section{An effective action for edge effects\label{sect:effective}}

\begin{figure}
\begin{center}
\includegraphics[height = 5.5cm]{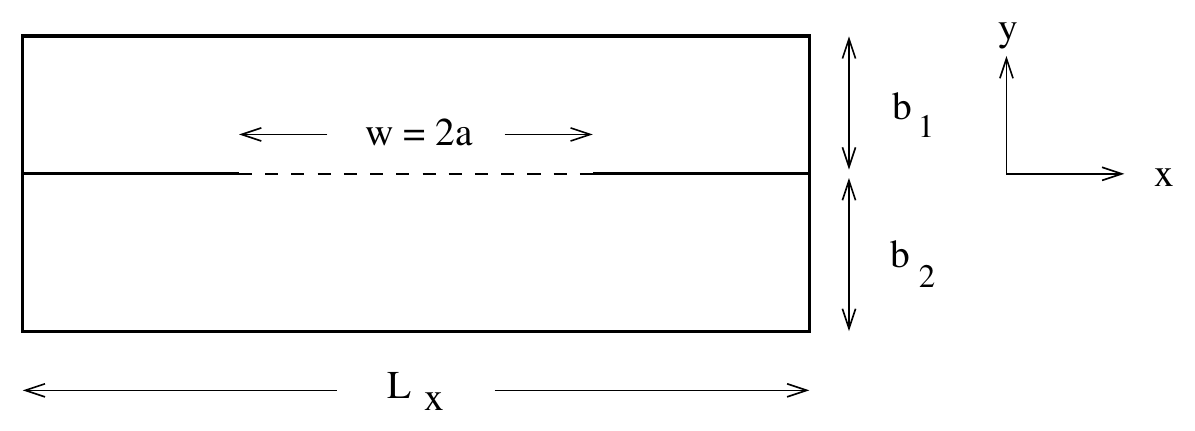}
\end{center}
\caption{A two-dimensional slice through the geometry.  Dirichlet boundary conditions
are imposed on the solid lines.  The gap between the plates (where the
non-local field theory lives) is indicated by a dashed line.  The
four-dimensional geometry also has a periodic spatial dimension of
size $L_z$ out of the page and a periodic Euclidean time dimension of size $\beta$.
\label{fig:FullGeometry}}
\end{figure}

We consider a free massless scalar field in four dimensions, with Dirichlet boundary
conditions imposed on an arrangement of plates.  The basic plate geometry we
will consider is shown in Fig.~\ref{fig:FullGeometry}.  Besides the two dimensions shown in the figure, the full geometry also
has a periodic spatial dimension of size $L_z$ and a periodic Euclidean
time dimension of size $\beta$.  For simplicity we will always have in
mind the limit $L_x,\,L_z \rightarrow \infty$, but as we are interested in
finite temperature we will keep $\beta$ fixed.

\begin{figure}
\begin{center}
\includegraphics[height = 4.5cm]{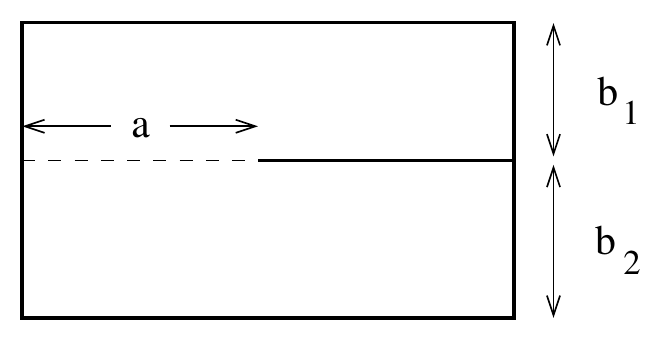}
\end{center}
\caption{The effective plate geometry for odd-parity modes.
\label{fig:HalfGeometry}}
\end{figure}

Starting from the geometry in Fig.~\ref{fig:FullGeometry}, but restricting to field
configurations which are odd under $x \rightarrow -x$, is equivalent to imposing a
Dirichlet boundary condition at $x = 0$.  That is, it corresponds to the effective
arrangement of plates shown in Fig.~\ref{fig:HalfGeometry}.

For reasons discussed below, we will focus on three special cases: 
\begin{itemize}
\item
a single plate with a slit, corresponding to $b_1,\,b_2 \rightarrow \infty$, $w = 2a$ fixed in
Fig.~\ref{fig:FullGeometry};
\item
perpendicular plates, corresponding to $b_1,\,b_2 \rightarrow \infty$, $a$ fixed in
Fig.~\ref{fig:HalfGeometry};
\item
parallel plates, corresponding to $a,\,b_1 \rightarrow \infty$, $b = b_2$ fixed in
Fig.~\ref{fig:HalfGeometry}.
\end{itemize}

The basic strategy, developed in \cite{Kabat:2010nm}, is to do the
Euclidean path integral in stages.  We first fix the value of the
field in the gap between the plates, setting $\phi = \phi_0$ on the
dashed line indicated in the figures, and subsequently integrate
over $\phi_0$.  In other words we write the Euclidean partition function as
\be
Z = \int {\cal D} \phi_0 \, \int_{\raisebox{-5pt}{$\phi \vert_{\rm gap} = \phi_0$}}
\hspace{-16mm} {\cal D} \phi \,\,\,\, e^{- \int d^4x \, {1 \over 2} \partial \phi \cdot \partial \phi}
\label{pathint2}
\ee
By integrating out the scalar field in the bulk regions (top and bottom) we obtain a non-local effective
action for $\phi_0$.  To perform the bulk path integral we set
$\phi = \phi_{\rm cl} + \delta \phi$
where $\delta \phi$ vanishes on all boundaries (including the gap), and $\Box \phi_{\rm cl} = 0$ subject to the boundary conditions
\be
\phi_{\rm cl} \rightarrow \left\lbrace \begin{array}{ll}
\phi_0 & \hbox{\rm in gap} \\
0 & \hbox{\rm elsewhere on boundary}
\end{array} \right.
\label{pathint4}
\ee
The action for $\delta\phi$ separates into top and bottom contributions leading to
\bea
\label{Z}
Z & =& \det{}^{-1/2}(-\Box_{\rm top}) \det{}^{-1/2}(-\Box_{\rm bottom}) \, \int {\cal D} \phi_0 \, e^{-S_0} \nonumber  \\
S_0 & = & \int d^4x \, {1 \over 2} \partial \phi_{\rm cl} \cdot \partial \phi_{\rm cl}
\eea
where $\Box_{\rm top},\,\Box_{\rm bottom}$ are the corresponding Laplacians.  Given the boundary conditions on $\delta \phi$, the bulk determinants are to be evaluated with Dirichlet boundary conditions everywhere, including the part of the boundary which corresponds to the gap. 

$\phi_{\rm cl}$ can be written in terms of $\phi_0$ and the Green's functions $G_{\rm top}$ and
$G_{\rm bottom}$.  These obey Dirichlet
boundary conditions and satisfy
$\Box G(x \vert x') = \delta^4(x-x')$ in the bulk regions. 
\begin{equation}
\label{Greens}
\phi_{\rm cl}(x) = \left\lbrace \begin{array}{ll}
\int d^3x' \, \phi_0(x') \, n \cdot \partial' G_{\rm top}(x \vert x') & \hbox{\rm on top} \\[8pt]
\int d^3x' \, \phi_0(x') \, n \cdot \partial' G_{\rm bottom}(x \vert x') & \hbox{\rm on bottom}
\end{array}\right.
\end{equation}
Here $n$ is an outward-pointing unit normal vector.  Integrating by
parts, the classical action in (\ref{Z}) becomes a surface term,
\begin{eqnarray}
\label{action1}
&& S_0 = \int d^{3}x \, \int d^{3}x' ~ {1 \over 2} \phi_0(x) \, (M_{\rm top} (x\vert x') + M_{\rm bottom} (x\vert x'))~ \phi_0 (x') \\[5pt]
\label{action2}
&& M(x\vert x') =n \cdot \partial \, n \cdot \partial' G (x \vert x')
\end{eqnarray}
The operator $M(x\vert x')$ is defined on the boundary between the bulk regions including the gap. 

The bulk determinants in (\ref{Z}) capture the Casimir energy
that would be present if there was no gap in the middle plate.  Corrections to this are
given by a non-local field theory that lives on the gap
separating the two regions. We can write a mode expansion for the fields
$\phi_0$ as
$\phi_0 ({x})
= \sum_\alpha c_\alpha u_\alpha ({ x})$ where $\left\{ u_\alpha ({x})
\right\}$ constitute a complete set of modes for functions which are nonzero
in the gap with the boundary condition that $u_\alpha ({x})\rightarrow 0$
as one approaches the edges. Integrating over $c_{\alpha}$ leads to a representation of 
the four-dimensional partition
function
\be
\label{Z4d}
Z_{\rm 4d} = \det{}^{-1/2}\big(-\Box_{\rm top}\big) \,
             \det{}^{-1/2}\big(-\Box_{\rm bottom}\big) \,
             \det{}^{-1/2}\big({\cal O}_{\rm top} + {\cal O}_{\rm bottom}\big)
\ee
where
\be
\label{matrix elements}
{\cal O}_{\alpha\beta} = \int_{\rm gap} d^{3}{x}~d^{3}{x'}~ u_\alpha (x) M (x\vert x') ~ u_\beta (x')
\ee
Because the mode functions $u_a(x)$ vanish outside the gap, the operators ${\cal O}$ are essentially the projected versions of $M(x|x')$ onto the gap, ${\cal O}=P M P$, where
$P$ is a projection operator
onto functions with support in the gap.  That is
\[
P f(x) = \left\lbrace\begin{array}{ll}
f(x) \quad & \hbox{\rm if $x \in {\rm gap}$} \\
0          & \hbox{\rm otherwise}
\end{array}\right.
\]

The explicit form of the operator $M(x\vert x')$ and its projected version ${\cal O}$ depends, in general,
on the arrangement of plates and gaps. For the geometries shown in Figs.~\ref{fig:FullGeometry} and
\ref{fig:HalfGeometry}, the non-local operators which appear in the effective action for $\phi_0$ are
\cite{Kabat:2010nm}
\bea
\label{operators}
&& {\cal O}_{\rm top} = P {\sqrt{-\nabla^2} \over \tanh \big(b_1 \sqrt{-\nabla^2}\big)} \, P \\
\nonumber
&& {\cal O}_{\rm bottom} = P {\sqrt{-\nabla^2} \over \tanh \big(b_2 \sqrt{-\nabla^2}\big)} \, P
\eea
Here $\nabla^2$ is the 3-dimensional Laplace operator defined on the
middle plate (including the gap) and $P$ is a projection operator
onto functions with support in the gap.\footnote{The asymptotic spectrum of such operators has recently been considered in
\cite{FractionalLaplacian}.}

At this stage it is convenient to make a Kaluza-Klein decomposition
along the two extra periodic directions.  This leads to a
representation of the four dimensional partition function in terms of a momentum integral and a sum over Matsubara frequencies.
\be
\label{Z4dKK}
\log Z_{\rm 4d} = L_z \int {dk \over 2\pi} \sum_{l = -\infty}^\infty
\log Z_{\rm 2d}\Big(\mu = \sqrt{k^2 + (2 \pi l / \beta)^2}\Big)
\ee
Here $Z_{\rm 2d}(\mu)$ is the two-dimensional partition function for a
scalar field of mass $\mu$ in the geometry shown in
Fig.~\ref{fig:FullGeometry} or \ref{fig:HalfGeometry}.

The representation (\ref{Z4dKK}) makes it apparent that in the high-temperature limit ($\beta \rightarrow 0$)
only the $l=0$ mode contributes and the problem reduces to a partition function in three dimensions.
Thus in the high-temperature limit the partition function is independent of $T$, and the free energy
is linear in $T$, independent of the geometry.\footnote{Strictly speaking this logic does not apply to ultraviolet
divergent parts of the partition function, and after renormalization divergent parts of the partition function can
make contributions to the free energy which grow as higher powers of $T$.  But since they are associated with UV
divergences, such contributions will necessarily be proportional to geometrical volumes or areas, and are
not conventionally regarded as part of the Casimir energy.  For an explicit example of this sort of behavior
see (\ref{4dhigh}).}  The behavior at low temperatures is more subtle
and will be considered in section \ref{sect:low}.

For the geometry of Fig.~\ref{fig:FullGeometry}, a complete set of  odd- and even-parity functions which vanish for $\vert x \vert \geqslant  a$ are
\bea
\label{OddModeFtns}
&& u^{\rm odd}_m = \left\lbrace
\begin{array}{cl}
(-1)^m {1\over \sqrt{a}} \sin \left(m \pi x / a\right) & \hbox{\rm for $-a \leqslant  x \leqslant a$}  \qquad m = 1,2,3,\ldots\\
0 & \hbox{\rm otherwise}
\end{array}\right. \\
\label{EvenModeFtns}
&& u^{\rm even}_p = \left\lbrace
\begin{array}{cl}
(-1)^{p + {1\over 2}} {1\over \sqrt{a}} \cos \left(p \pi x / a\right) & \hbox{\rm for $-a \leqslant  x \leqslant a$}  \qquad p = {1 \over 2},{3 \over 2},{5 \over 2},\ldots\\
0 & \hbox{\rm otherwise}
\end{array}\right.
\eea
Matrix elements of the operators (\ref{operators}) can be
evaluated in this basis as in (\ref{matrix elements}).  For the operator (denoting $d_x = {d \over dx}$)
\[
{\cal O} = P {\sqrt{-d_x^2 + \mu^2} \over \tanh \big(b \sqrt{-d_x^2 + \mu^2}\big)} \, P
\]
we have the matrix elements
\bea
{\cal O}^{\rm odd}_{mn} & = & {2 a \over \pi} \int_{-\infty}^\infty dk \, \sin^2(ka) \, M(k) \, {m\pi \over k^2 a^2 - m^2 \pi^2} \,
{n\pi \over k^2 a^2 - n^2 \pi^2} \\
{\cal O}^{\rm even}_{pq} & = & {2 a \over \pi} \int_{-\infty}^\infty dk \, \cos^2(ka) \, M(k) \, {p\pi \over k^2 a^2 - p^2 \pi^2} \,
{q\pi \over k^2 a^2 - q^2 \pi^2} 
\eea
where $m,n=1,2,\cdots$,~$p,q=1/2,3/2,\cdots$, and  $M(k) = {\sqrt{k^2 + \mu^2} \over \tanh \big(b \sqrt{k^2 + \mu^2}\big)}$ has the useful representation
\[
M(k) = {1 \over b} + {2 \over b} \sum_{j = 1}^\infty {k^2 + \mu^2 \over k^2 + \mu^2 + {j^2 \pi^2 \over b^2}}\,.
\]
As discussed in \cite{Kabat:2010nm}, by contour deformation the matrix
elements can be decomposed into ``direct'' and ``diffractive''
contributions.\footnote{In \cite{Kabat:2010nm} these were referred to
as ``pole'' and ``cut'' contributions, respectively.}  For the odd matrix elements
\bea
\label{Odd}
&& {\cal O}^{\rm odd}_{mn} = {\cal O}_{mn}^{\rm direct} + {\cal O}_{mn}^{\rm diffractive} \\
\label{OddDirect}
&& {\cal O}_{mn}^{\rm direct} = {\sqrt{(m \pi / {a})^2 + \mu^2} \over \tanh \big(b \sqrt{(m \pi / {a})^2 + \mu^2}\big)} \, \delta_{mn} \\
\nonumber
&& {\cal O}_{mn}^{\rm diffractive} = - 2 a b^2 \sum_{j = 1}^\infty
\left(1 - \exp\Big(-{2 a \over b}\sqrt{j^2 \pi^2 + \mu^2 b^2}\Big)\right)
{j^2 \pi^2 \over \sqrt{j^2 \pi^2 + \mu^2 b^2}} \\
\label{OddDiffractive}
&& \qquad \qquad \qquad {m \pi \over (m \pi b)^2 + (j \pi a)^2 + (\mu a b)^2} \,\,
                        {n \pi \over (n \pi b)^2 + (j \pi a)^2 + (\mu a b)^2}
\eea
Likewise for the even matrix elements
\bea
\label{Even}
&& {\cal O}^{\rm even}_{pq} = {\cal O}_{pq}^{\rm direct} + {\cal O}_{pq}^{\rm diffractive} \\
\label{EvenDirect}
&& {\cal O}_{pq}^{\rm direct} = {\sqrt{(p \pi / {a})^2 + \mu^2} \over \tanh \big(b \sqrt{(p \pi / {a})^2 + \mu^2}\big)} \, \delta_{pq} \\
\nonumber
&&{\cal O}_{pq}^{\rm diffractive} = - 2 a b^2 \sum_{j = 1}^\infty
\left(1 + \exp\Big(-{2 a \over b}\sqrt{j^2 \pi^2 + \mu^2 b^2}\Big)\right)
{j^2 \pi^2 \over \sqrt{j^2 \pi^2 + \mu^2 b^2}} \\
\label{EvenDiffractive}
&& \qquad \qquad \qquad {p \pi \over (p \pi b)^2 + (j \pi a)^2 + (\mu a b)^2} \,\,
                        {q \pi \over (q \pi b)^2 + (j \pi a)^2 + (\mu a b)^2}
\eea
(Aside from the allowed values of the indices, the only difference between odd and even parity is the sign
in front of the exponential in the diffractive term.)  Finally, to study the perpendicular plate geometry of Fig.~\ref{fig:HalfGeometry}, we only need to keep
the odd-parity modes (\ref{OddModeFtns}).  Thus the matrix elements for perpendicular plates are exactly those given in (\ref{Odd}) -- (\ref{OddDiffractive}).

The direct contribution takes into account wave propagation directly
across the gap.  Note that it is diagonal in the basis we are using.
Mathematically ${\cal O}^{\rm direct}$ is simply the operator
${\sqrt{-d_x^2 + \mu^2} \over \tanh \big(b \sqrt{-d_x^2 +
\mu^2}\big)}$, defined with Dirichlet boundary conditions at $x = -a$
and $x = a$.  Corrections to this, which incorporate diffraction of
waves through the gap, are encoded in ${\cal O}^{\rm diffractive}$.

The approach developed in \cite{Kabat:2010nm} was to treat diffraction
as a small perturbation.  Taking the log of (\ref{Z4d}) and expanding
in powers of ${\cal O}^{\rm diffractive}$, the 4d free energy naturally
decomposes into bulk, direct and diffractive contributions.
\bea
\label{bulk}
-\log Z_{\rm bulk} &=& {1 \over 2} {\rm Tr} \log \big(-\Box_{\rm top}\big)
                               + {1 \over 2} {\rm Tr} \log \big(-\Box_{\rm bottom}\big) \\
\label{direct}
-\log Z_{\rm direct} &=& {1 \over 2} {\rm Tr} \log \big({\cal O}_{\rm top}^{\rm direct}
                                                                + {\cal O}_{\rm bottom}^{\rm direct}\big) \\
\label{diffractive}
-\log Z_{\rm diffractive} &= &{1 \over 2} {\rm Tr} \left[ 
\left({\cal O}^{\rm direct}\right)^{-1} {\cal O}^{\rm diffractive} \right] \\
\nonumber
&&-{1 \over 4} {\rm Tr }\left[ 
\left({\cal O}^{\rm direct}\right)^{-1}
{\cal O}^{\rm diffractive}
\left({\cal O}^{\rm direct}\right)^{-1}
 {\cal O}^{\rm diffractive} \right] \\[3pt]
 \nonumber
 && + \cdots \\
 \label{expansion}
 &=& -\log Z^{(1)}_{\rm diffractive} -\log Z^{(2)}_{\rm diffractive} + \cdots
\eea
where in (\ref{diffractive})
\bea
&&{\cal O}^{\rm direct} = {\cal O}_{\rm top}^{\rm direct} + {\cal O}_{\rm bottom}^{\rm direct} \nonumber\\
&&{\cal O}^{\rm diffractive} = {\cal O}_{\rm top}^{\rm diffractive} + {\cal O}_{\rm bottom}^{\rm diffractive} 
\eea

The bulk and direct contributions (\ref{bulk}), (\ref{direct}) are basically Bose partition functions and can be
calculated analytically.  The relevant calculations are summarized in appendix \ref{sect:thermo}.
Our main interest in the next section will be diffractive effects.

\section{Thermal free energy: low temperature limit\label{sect:low}}

In this section we study the behavior of the partition function (\ref{Z4dKK}) at low temperatures.  Applying Poisson resummation to (\ref{Z4dKK}) gives
\be
\label{Z4dthermal1}
\log Z_{\rm 4d} = \sum_{l = -\infty}^\infty \beta L_z \int {dk \over 2\pi} {d\omega \over 2\pi}
e^{-i \beta \omega l} \log Z_{\rm 2d}\big(\mu = \sqrt{k^2 + \omega^2}\big)
\ee
The $l = 0$ term is proportional to $\beta$.  It gives the Casimir
energy at zero temperature that was studied in \cite{Kabat:2010nm}.
Thermal corrections to this are given by
\be
\label{Z4dthermal}
\log Z_{\rm 4d,T} = {\beta L_z \over \pi} \sum_{l = 1}^\infty \int_0^\infty \mu d\mu J_0(\beta l \mu)
\log Z_{\rm 2d}(\mu)
\ee
where we set $\omega = \mu \cos \theta$, $k = \mu \sin \theta$ and integrated over $\theta$.

It is clear that the behavior of (\ref{Z4dthermal}) at low temperature, $\beta \rightarrow \infty$, is related to the
behavior of $\log Z_{\rm 2d}$ as $\mu \rightarrow 0$.  For instance if $\log Z_{\rm 2d}(\mu)$ is analytic as a function
of $\mu^2$ along the positive real $\mu^2$ axis then the 4d free energy will vanish exponentially at low temperature.\footnote{To see this return to the
representation (\ref{Z4dthermal1}).  Note that analyticity of $\log Z_{\rm 2d}$ for positive real $\mu^2$ implies analyticity for positive real $\omega^2$.
Then the integrand in (\ref{Z4dthermal1}) is analytic along the real $\omega$ axis and the $\omega$ contour
of integration can be deformed into the upper or lower half plane.  This shows that terms with $l \not= 0$ are exponentially small.}
On the other hand, assuming that $\log Z_{\rm
2d}$ does not diverge for large $\mu$, we can use \footnote{One can make this well-defined by
inserting a convergence factor $e^{-\alpha x}$ and using
\[
\int_0^\infty dx \, e^{-\alpha x} J_0(\beta x) x^{\nu - 1} = {\Gamma(\nu) \over \alpha^\nu}
F\left({\nu \over 2},{\nu + 1 \over 2},1,-{\beta^2 \over \alpha^2}\right)\,.
\]
The final answer is independent of $\alpha$ as $\beta \rightarrow \infty$ and yields (\ref{BesselIntegral}).}
\be
\label{BesselIntegral}
\int_0^\infty dx \, J_0(\beta x) x^{\nu - 1} = - {2^\nu \Gamma(\nu/2) \over \nu \Gamma(-\nu/2)} \, {1 \over \beta^\nu}
\ee
So power-law behavior of the 2d free energy as $\mu \rightarrow 0$,
$\log Z_{\rm 2d} \sim \mu^{\nu - 2}$, will in general lead to power-law behavior
of the 4d free energy at low temperature, $\log Z_{\rm 4d} \sim
T^\nu$.  (In accord with our analyticity arguments, the coefficient of $T^\nu$ vanishes for $\nu = 2,4,6,\ldots$)

For future use it is convenient to define $f(\nu) = - {2^\nu
\Gamma(\nu/2) \over \nu \Gamma(-\nu/2)}$.  Differentiating
(\ref{BesselIntegral}) with respect to $\nu$ gives the useful
identities
\bea
\label{BesselInt2}
&&\int_0^\infty dx \, J_0(\beta x) x^{\nu - 1} = f(\nu) \, {1 \over \beta^\nu} \\
\label{BesselInt3}
&&\int_0^\infty dx \, J_0(\beta x) x^{\nu - 1} \log x = - f(\nu) \, {1 \over \beta^\nu} \log \beta + f'(\nu) {1 \over \beta^\nu} \\
\label{BesselInt4}
&&\int_0^\infty dx \, J_0(\beta x) x^{\nu - 1} (\log x)^2 = f(\nu) \, {1 \over \beta^\nu} (\log \beta)^2
- 2f'(\nu) {1 \over \beta^\nu} \log \beta + f''(\nu) {1 \over \beta^\nu} \qquad
\eea

We will evaluate thermal contributions to the free energy using
the representation (\ref{Z4dthermal}).  If the geometric
parameters $a,b_1,b_2$ are held fixed then, from (\ref{OddDirect}) -- (\ref{EvenDiffractive}),
all matrix elements are analytic in $\mu^2$ about $\mu = 0$.  The 2d partition function inherits this analyticity, which means that at low temperatures the 4d free energy
is exponentially suppressed.  As a result, we proceed to study three special cases which have interesting
power-law behavior at low temperature:
\begin{itemize}
\item
perpendicular plates,
\item
a slit geometry,
\item
parallel plates.
\end{itemize}

\subsection{Perpendicular plates\label{sect:perp}}

\begin{figure}[h]
\begin{center}
\includegraphics[height = 4.5cm]{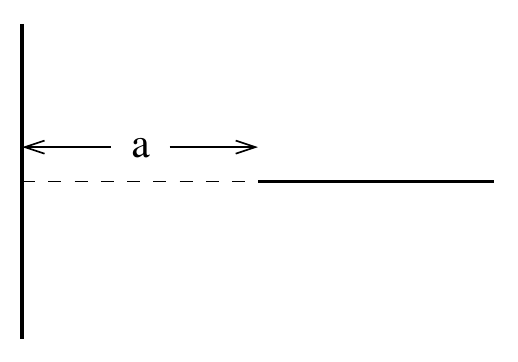}
\end{center}
\caption{Perpendicular plates.  The dashed line indicates the gap
between the plates.  There is also a periodic spatial dimension of
size $L_z \rightarrow \infty$ pointing out of the page and a periodic
Euclidean time dimension of size $\beta$.
\label{fig:perp}}
\end{figure}

\begin{figure}[h]
\begin{center}
\includegraphics[height = 5cm]{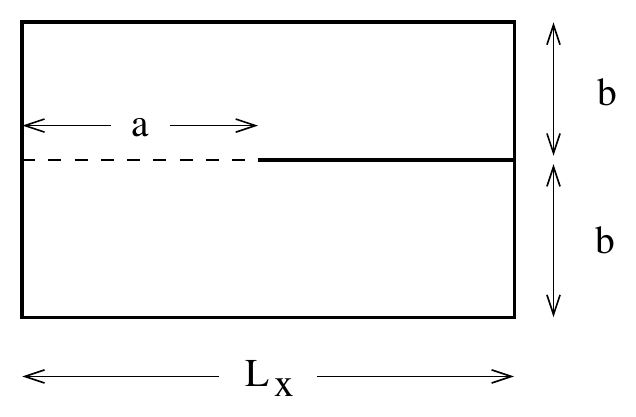}
\end{center}
\caption{Regulated geometry for perpendicular plates.
\label{fig:perp2}}
\end{figure}

In this section we study the low temperature behavior of the free energy
for perpendicular plates.  The geometry of interest is shown in Fig.~\ref{fig:perp}.
However to regulate IR divergences we actually work with the geometry of Fig.~\ref{fig:perp2}
in the limit $b,\,L_x \rightarrow \infty$.

\noindent
There are three contributions to the thermal free energy.

\noindent
{\em Bulk contribution} \\
The bulk contribution (\ref{bulk}) from the regions above and below the middle plate is that of an ideal Bose gas.
This is worked out in (\ref{4dhigh}).  Including surface contributions associated with the Dirichlet boundary conditions,
the free energy is
\be
\label{freegas}
F^{\rm bulk}_{\rm top} = F^{\rm bulk}_{\rm bottom} =  - {\zeta(4) \over \pi^2} b L_x L_z T^4 + {\zeta(3) \over 4 \pi} (L_x +b) L_z T^3 - {\zeta(2) \over 4\pi} L_z T^2  
\ee
To isolate the thermal Casimir energy associated with the gap in the middle plate we proceed
as follows.  First we subtract the free energy of a ``big box'' of volume $2b \times L_x \times L_z$
without any middle plate.  This is given by
\be
\label{box}
F_{\rm box} = -{\zeta(4) \over \pi^2} 2 b L_x L_z T^4 + {\zeta(3) \over 4 \pi} (L_x + 2b) L_z T^3
- {\zeta(2) \over 4\pi} L_z T^2
\ee
Next we subtract the thermal self-energy of the middle plate itself, as well as the thermal self-energy associated
with the ``$\dashv\,$'' shaped junction on the right side of Fig.~\ref{fig:perp2}.  These are given by
\be
F_{\rm self}
= {\zeta(3) \over 4 \pi} (L_x - a) L_z T^3 - {\zeta(2) \over 8 \pi} L_z T^2
\ee
Thus the bulk contribution to the thermal Casimir free energy for perpendicular plates is
\be
\label{bulkperp}
F^{\rm bulk}_{\perp, T} = F^{\rm bulk}_{\rm top} + F^{\rm bulk}_{\rm bottom} - F_{\rm box} - F_{\rm self}
= {\zeta(3) \over 4\pi} L_z a T^3 - {\zeta(2) \over 8 \pi} L_z T^2
\ee
Eq.\ (\ref{bulkperp}) provides the leading low temperature behavior of the Casimir energy and it agrees with the results on the thermal Casimir force found in \cite{KlingGies,GiesWeber}.

\medskip
\noindent
{\em Direct contribution} \\
To evaluate the direct contribution to the free energy (\ref{direct}),
note from (\ref{OddDirect}) that as $b \rightarrow \infty$ the direct matrix elements are given by
\be
\label{PerpOdirect}
{\cal O}_{mn}^{\rm direct} = \sqrt{(m \pi / a)^2 + \mu^2} \, \delta_{mn}
\ee
Thus the direct contribution to the free energy can be identified with half the free energy of an ideal gas in $2+1$ dimensions,
where the gas occupies the region corresponding to the gap.  This free energy is worked out in appendix \ref{sect:thermo} equation (\ref{3dlow}).  We find that
\be
\label{directperp}
F^{\rm direct}_{\perp,T}  = - {L_z T \over {2 a}} \sum _{m,n=1}^{\infty} {{m} \over {n}}K_1({{mn \pi} /{aT}})
\ee
At low temperatures, $aT<<1$, the direct contribution to the thermal free energy is exponentially suppressed since the thermal wavelength does not fit in the gap.\footnote{This also follows from the fact that the matrix elements (\ref{PerpOdirect}) are analytic in $\mu^2$ about $\mu = 0$.}

\medskip
\noindent
{\em First diffractive contribution} \\
To evaluate the diffractive contribution to the free energy
(\ref{diffractive}) we need to study the operator ${\cal O}^{\rm
diffractive}_{\perp}$.  As $b \rightarrow \infty$ we can replace the sum in
(\ref{OddDiffractive}) with an integral to obtain
\be
\label{PerpOdiffractive}
{\cal O}_{\perp,mn}^{\rm diffractive} = - {2 \over \pi a^3} \int_0^\infty dk \,
\left(1 - e^{-2 a \sqrt{k^2 + \mu^2}}\right)
{k^2 \over \sqrt{k^2 + \mu^2}} \,\,
{m \pi \over (m \pi / a)^2 + k^2 + \mu^2} \,\,
{n \pi \over (n \pi / a)^2 + k^2 + \mu^2}
\ee
Using (\ref{PerpOdirect}) and (\ref{PerpOdiffractive}) in
(\ref{diffractive}), at first order in perturbation theory the
diffractive contribution to the 2d partition function is,
with $x = a \sqrt{k^2 + \mu^2}$,
\be
\label{PerpZ2d}
- \log Z_{\rm 2d}^{\rm diffractive} = - {1 \over \pi} \sum_{n = 1}^\infty {n^2 \pi^2 \over \sqrt{n^2 \pi^2 + \mu^2 a^2}} \,
\int_{\mu a}^\infty dx \sqrt{x^2 - \mu^2 a^2} \, \big(1 - e^{-2x}\big) {1 \over \big(x^2 + n^2 \pi^2\big)^2}
\ee
At this point we need to determine the non-analytic behavior as $\mu
\rightarrow 0$ of the integral
\be
\label{integralI}
I = \int_{\mu a}^\infty dx \sqrt{x^2 - \mu^2 a^2}  \, \left[\big(1 - e^{-2x}\big) {1 \over \big(x^2 + n^2 \pi^2\big)^2}\right]
\ee
To obtain this we split the region of integration in two, introducing an intermediate scale $x_0$ with
$\mu a \ll x_0 \ll 1$. We evaluate the integral (\ref{integralI}) in the region $\mu a < x < x_0$ by expanding the quantity in square brackets in powers of
$x$ and integrating term-by-term.  Only even powers of $x$ in this expansion give contributions which are non-analytic
in $\mu^2$.  Similarly, we evaluate the integral in the region $x_0 < x < \infty$ by expanding
$\sqrt{x^2-\mu^2 a^2}$ in powers of $\mu^2$ and integrating term-by-term.  This of course gives a contribution which is analytic in
$\mu^2$.  One can check that, order by order, the final result does not depend on the intermediate scale $x_0$.
This procedure gives
\bea
\label{nonanalytic}
I =&& - {1 \over 4 n^4 \pi^4} (\mu a)^4 \log \mu a + \left({1 \over 4 n^6 \pi^6} - {1 \over 24 n^4 \pi^4}\right)
(\mu a)^6 \log \mu a + \cdots \nonumber \\
&& + \, (\hbox{\rm terms analytic in $\mu^2$})
\eea
Substituting this in (\ref{PerpZ2d}) and evaluating the sum on $n$
gives the non-analytic behavior of the 2d partition function.
\be
\label{firstdiffr}
- \log Z_{2d}^{\rm diffractive} = {\zeta(3) \over 4 \pi^4} (\mu a)^4 \log \mu a + \left({\zeta(3) \over 24 \pi^4}
- {3 \zeta(5) \over 8 \pi^6}\right) (\mu a)^6 \log \mu a + \cdots
\ee
From (\ref{Z4dthermal}) the leading low temperature behavior of the 4d
thermal free energy is then
\[
F_{\perp, T}^{\rm diffractive} = {L_z \zeta(3) \over 4 \pi^5 a^2} \sum_{l = 1}^\infty \int_0^\infty d(\mu a)
J_0(\beta l \mu) (\mu a)^5 \log(\mu a)
\]
This integral is evaluated using (\ref{BesselInt3}).  In the case at hand
$f(6) = 0$ and $f'(6) = - 64$.  Doing the sum on $l$, the first diffractive
correction to the free energy is
\be
\label{F4dPerp}
F_{\perp, T}^{(1) \rm diffractive} = - {16 \pi \zeta(3) \over 945} L_z a^4 T^6 + {\cal O}(T^8)
\ee

\medskip
\noindent
{\em Higher diffractive contributions} \\
In \cite{Kabat:2010nm} we studied higher order terms in the expansion (\ref{expansion}).  We found that the $n^{th}$ order diffractive
contribution to $-\log Z_{\rm 2d}$ for perpendicular plates is of the form 
\be
-\log Z_{\rm 2d}^{(n){\rm diffractive}} =
- {2^{n-1}\over n} \int_1^\infty \prod^n_{i=1} dy_i \sqrt{y_i^2-1} \left(1- e^{-2 \mu a y_i}\right) T(\mu a)\ast T(\mu a)\ast \cdots
\ast T(\mu a)
\ee
where
\be
T\ast T \ast \cdots \ast T = T(\mu a, y_1, y_2) T(\mu a, y_2, y_3) \cdots
T(\mu a, y_n, y_1)\nonumber
\ee
and 
\be
T (\mu a, y, z) =  {\mu^2 a^2 \over \pi} \sum_{r=1,2,\cdots}
{  r^2 \pi^2 \over \sqrt{r^2\pi^2 + \mu^2 a^2}~ ( r^2 \pi^2 + \mu^2 a^2 y^2 )~( r^2 \pi^2 + \mu^2 a^2 z^2)}
\label{slit-diff6}
\ee
Using the change of variables $x_i=\mu a y_i$ and analyzing each of the $x_i$ integrals as outlined below (\ref{integralI}), we find that
the small $\mu$ behavior of the 2d partition function is characterized by analytic and non-analytic terms of the form
\bea
\label{n-diffr}
- \log Z_{\rm 2d}^{(n){\rm diffractive}} &=& \sum_{r=1}^{n} \sum_{l = 0}^\infty A_{rl} (\mu a )^{4r +2l} (\log \mu a)^ r \\
&+& (\hbox{\rm terms analytic in $\mu^2$}) \nonumber
\eea
The second order diffractive contribution to the $(\mu a)^4 \log({\mu a})$ term is $0.00031 ~(\mu a)^4 \log\mu a$. This is  a $10\%$ correction compared to the first order term ${\zeta(3) \over 4 \pi^4} (\mu a)^4 \log \mu a$ in (\ref{firstdiffr}). Higher order effects are much smaller.

Logarithms of the temperature arise at higher orders in perturbation theory.  Indeed, using relations similar to (\ref{BesselInt4}), but applied to higher $\log$-terms, we find that the $n^{th}$ order diffractive term in the perturbative expansion contributes to the thermal free energy new log-terms of the form $T^{4n+2+2l} (\log T)^{n-1},~~l=0,1,\cdots$. The first $\log T$ term is of order $T^{10} \log T$ and can be neglected at low $T$.

\medskip
\noindent
{\em Summary} \\ 
Collecting our results, the low temperature behavior of the free energy for perpendicular plates, up to first order in diffractive effects, is
\be
\label{PerpSummary}
F_{\perp, T} = - {\zeta(2) \over 8 \pi} L_z T^2 + {\zeta(3) \over 4\pi} L_z a T^3 - {16 \pi \zeta(3) \over 945} L_z a^4 T^6 + {\cal O}(T^8)
\ee
The two leading terms come from the bulk determinants.  They have a simple
physical interpretation.  At low temperatures ($aT \ll 1$) the
thermal wavelength is larger than the size of the gap.  As a result
the field does not see the gap and behaves as though a Dirichlet
boundary condition had been imposed there.  So the thermal
renormalization of the tension associated with a Dirichlet boundary
and a ``$\,\vdash$'' shaped junction
also applies to the gap.  This effect can be thought of as an excluded area effect, and is responsible for the leading
low temperature behavior of the Casimir energy.  Diffractive effects
are subleading, beginning at ${\cal O}(T^6)$, while the direct
contribution from the theory in the gap is exponentially suppressed.

\subsection{Slit geometry\label{sect:slit}}
\begin{figure}[h]
\begin{center}
\includegraphics[width = 10.8cm]{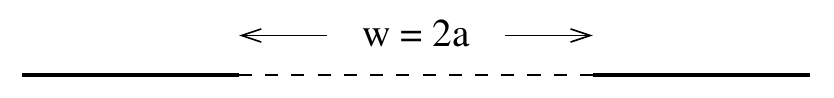}
\end{center}
\caption{Slit geometry.  The dashed line indicates the gap
between the plates.  There is also a periodic spatial dimension of
size $L_z \rightarrow \infty$ pointing out of the page and a periodic
Euclidean time dimension of size $\beta$.
\label{fig:slit}}
\end{figure}

In this section we study the low temperature behavior of the free energy for a slit of width $w=2a$. The corresponding geometry is shown in Fig.~\ref{fig:slit}.

\noindent
There are again three contributions to the thermal free energy. 

\noindent
{\em Bulk contribution} \\
The bulk contribution for the slit is very similar to the one found for the perpendicular plates. The contribution from the regions above and below the middle plate is that of an ideal Bose gas as in (\ref{freegas}). In isolating the thermal Casimir energy associated with the slit, we subtract the free energy of the ``big box" as given in (\ref{box}) and the self energy of the middle plate which is
\be
F_{\rm self}
= {\zeta(3) \over 4 \pi} (L_x - w) L_z T^3 - {\zeta(2) \over 4 \pi} L_z T^2
\ee
The final bulk contribution to the free energy is
\be
F^{\rm bulk}_{\perp, T} = F^{\rm bulk}_{\rm top} + F^{\rm bulk}_{\rm bottom} - F_{\rm box} - F_{\rm self}
= {\zeta(3) \over 4\pi} L_z w T^3
\ee

\medskip
\noindent
{\em Direct contribution} \\
To evaluate the direct contribution to the free energy (\ref{direct}),
note from (\ref{OddDirect}) and (\ref{EvenDirect}), that as $b \rightarrow \infty$ the direct matrix elements are given by
\be
\label{SlitOdirect}
{\cal O}_{ll'}^{\rm direct} = \sqrt{(l \pi / w)^2 + \mu^2} \, \delta_{ll'}
\ee
where both odd (\ref{OddDirect}) and even terms (\ref{EvenDirect}) have been included. The direct contribution to the free energy is given by (\ref{directperp}), where $a \rightarrow w$, and it is exponentially suppressed as expected.

\medskip
\noindent
{\em First diffractive contribution} \\
To evaluate the diffractive contribution to the free energy
(\ref{diffractive}) we need to study the operator ${\cal O}^{\rm
diffractive}_{\rm slit}$.  As $b \rightarrow \infty$ we can replace the sum in
(\ref{OddDiffractive}) and (\ref{EvenDiffractive}) with an integral to obtain
\be
\label{SlitOdiffractive}
{\cal O}_{{\rm slit},ll'}^{\rm diffractive} = - {4 \over \pi w^3} \int_0^\infty dk \,
\left[1 - (-1)^l e^{-w \sqrt{k^2 + \mu^2}}\right]
{k^2 \over \sqrt{k^2 + \mu^2}} \,\,
{l \pi \over (l \pi / w)^2 + k^2 + \mu^2} \,\,
{l' \pi \over (l' \pi / w)^2 + k^2 + \mu^2}
\ee
Using (\ref{SlitOdirect}) and (\ref{SlitOdiffractive}) in
(\ref{diffractive}), at first order in perturbation theory, the
diffractive contribution to the 2d partition function is
\bea
\label{SlitZ2d}
&&- \log Z_{\rm 2d}^{\rm diffractive}  = - \log Z_{\rm 2d}^{\rm odd, diffractive}- \log Z_{\rm 2d}^{\rm even, diffractive} \\
&=&- {4 \over {\pi w^3}} \sum_{l = 1}^\infty {l^2 \pi^2 \over \sqrt{(l \pi/w)^2 + \mu^2 }} 
 \int_0^\infty dk \,
\left[1 - (-1)^l e^{-w \sqrt{k^2 + \mu^2}}\right]
{k^2 \over \sqrt{k^2 + \mu^2}} \,
{1 \over {[(l \pi / w)^2 + k^2 + \mu^2]^2} }   \nonumber
\eea
where $ - \log Z_{\rm 2d}^{\rm odd, diffractive}$ accounts for the contribution of the odd modes $l=2m$ and $ - \log Z_{\rm 2d}^{\rm even, diffractive}$ accounts for the contribution of the even modes $l=2p+1$. Using $w=2a$ and changing variables to $x=a \sqrt{k^2+\mu^2}$ we find that $ - \log Z_{\rm 2d}^{\rm odd, diffractive}$ is identical to the expression (\ref{PerpZ2d}) for the perpendicular plates. As we saw in section 3.1, this produces a diffractive correction to the thermal free energy of order $T^6$ at low temperatures, namely
\be
\label{F4dSlit odd}
F_{\rm slit, T}^{(1) \rm odd, diffractive} = - {{16 \pi \zeta(3)} \over {945 }} L_z a^4 T^6 + {\cal O}(T^8)
\ee
Next we focus on the contribution of the even modes.
\be
\label{SlitevenZ2d}
- \log Z_{\rm 2d}^{\rm even, diffractive} = - {1 \over \pi} \sum_{p = 1/2, 3/2,\cdots} {p^2 \pi^2 \over \sqrt{p^2 \pi^2 + \mu^2 a^2}} \,
\int_{\mu a}^\infty dx \sqrt{x^2 - \mu^2 a^2} \, \big(1 + e^{-2x}\big) {1 \over \big(x^2 + p^2 \pi^2\big)^2}
\ee
At this point we need to determine the non-analytic behavior as $\mu
\rightarrow 0$ of the integral
\be
\label{integralI'}
I'= \int_{\mu a}^\infty dx \sqrt{x^2 - \mu^2 a^2}  \, \left[\big(1 + e^{-2x}\big) {1 \over \big(x^2 + p^2 \pi^2\big)^2}\right]
\ee
Following the same analysis we did for (\ref{integralI}) in the case of the perpendicular plates we find 
\bea
\label{nonanalyticI'}
I' =&&  {1 \over  p^4 \pi^4} (\mu a)^2 \log \mu a + \left({1 \over 4 p^4 \pi^4} - {1 \over 2 p^6 \pi^6}\right)
(\mu a)^4 \log \mu a + \cdots \nonumber \\
&& + \, (\hbox{\rm terms analytic in $\mu^2$})
\eea
Substituting this in (\ref{SlitevenZ2d}) and evaluating the sum on $p$
gives the non-analytic behavior of the even 2d partition function.
\be
\label{firstevendiffr}
- \log Z_{2d}^{\rm even, diffractive} = -{{7 \zeta(3)} \over  \pi^4} (\mu a)^2 \log \mu a - \left({{7\zeta(3)} \over 4 \pi^4}
- {31 \zeta(5) \over  \pi^6}\right) (\mu a)^4 \log \mu a + \cdots
\ee
Using (\ref{Z4dthermal}) and (\ref{BesselInt3}) we find that the leading low temperature behavior of the even 4d
thermal free energy is 
\be
\label{F4dSlit even}
F_{\rm slit, T}^{(1) \rm even, diffractive} = - {{14 \zeta(3)} \over {45 \pi}} L_z a^2 T^4 + {\cal O}(T^6)
\ee
Comparing (\ref{F4dSlit odd}) and (\ref{F4dSlit even}) we see that the even modes dominate the diffractive contribution to the free energy at low temperatures. So, for a slit of width $w$,
\bea
F_{\rm slit, T}^{(1) \rm diffractive} &=& F_{\rm slit, T}^{(1) \rm odd, diffractive} + F_{\rm slit, T}^{(1) \rm even, diffractive} \nonumber \\
\label{F4dSlit}
& = &- {{7 \zeta(3)} \over {90 \pi}} L_z w^2 T^4 + {\cal O}(T^6)
\eea

\medskip
\noindent
{\em Higher diffractive contributions} \\
In \cite{Kabat:2010nm} we studied higher order terms in the expansion (\ref{expansion}).  We found that the even $n^{th}$ order diffractive
contribution to $-\log Z_{\rm 2d}$ for a slit of width $w=2a$ is of the form 
\be
-\log Z_{\rm 2d}^{{\rm even},(n)} =
- {2^{n-1}\over n} \int_1^\infty \prod^n_{i=1} dy_i \sqrt{y_i^2-1} \left(1+ e^{-2 \mu a y_i}\right) S(\mu a)\ast S(\mu a)\ast \cdots
\ast S(\mu a)
\ee
where
\be
S\ast S \ast \cdots \ast S = S(\mu a, y_1, y_2) S(\mu a, y_2, y_3) \cdots
S(\mu a, y_n, y_1)\nonumber
\ee
and 
\be
S (\mu a, y, z) =  {\mu^2 a^2 \over \pi} \sum_{r=1/2,3/2,\cdots}
{  r^2 \pi^2 \over \sqrt{r^2\pi^2 + \mu^2 a^2}~ ( r^2 \pi^2 + \mu^2 a^2 y^2 )~( r^2 \pi^2 + \mu^2 a^2 z^2)}
\label{slit-diff7}
\ee
Using the change of variables $x_i=\mu a y_i$ and analyzing each of the $x_i$ integrals as outlined below (\ref{integralI}), we find that the small $\mu$ behavior of the 2d partition function is 
\bea
\label{n-diffr/slit}
- \log Z_{\rm 2d}^{{\rm even},(n)} &=&
 \sum_{r=1}^{n} \sum_{l = 0}^\infty B_{rl} (\mu a )^{2r +2l} (\log a \mu )^ r \\
&& + (\hbox{\rm terms analytic in $\mu^2$})  \nonumber
\eea
The second order diffractive contribution to the $(\mu a)^2 \log({\mu a})$ term is $-0.00526 ~(\mu a)^2 \log\mu a$. This is a $6\%$ correction compared to the first order term $-{{7 \zeta(3)} \over  \pi^4} (\mu a)^2 \log \mu a$ in (\ref{firstevendiffr}). Higher order effects are much smaller.

As explained earlier in the case of the higher diffractive contributions for the perpendicular plates,  each $n^{th}$ order diffractive term in the perturbative expansion contributes new log-terms to the thermal free energy of the type $T^{2n+2+2l} (\log T)^{n-1},~~l=0,1,\cdots$. The first $\log T$ term is of order $T^{6} \log T$ and can be neglected at low $T$.

\medskip
\noindent
{\em Summary} \\ 
Collecting our results, the low temperature behavior of the free energy for a slit, up to first order in diffractive effects, is
\be
\label{SlitSummary}
F_{\rm slit, T} =  {\zeta(3) \over 4\pi} L_z w T^3 - {7  \zeta(3) \over {90 \pi}} L_z w^2 T^4 + {\cal O}(T^6)
\ee
The leading contribution comes from the bulk determinants.  The direct contribution from the theory in the gap is exponentially suppressed while the diffractive contribution is subleading, beginning at ${\cal O}(T^4)$.

\subsection{Parallel plates\label{sect:parallel}}

\begin{figure}[h]
\begin{center}
\includegraphics[width = 9.45cm]{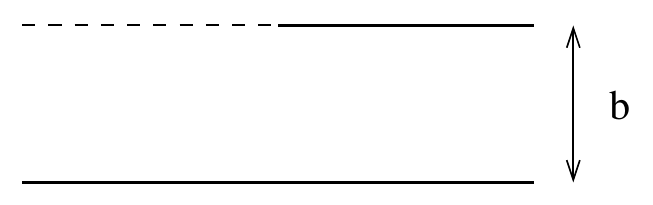}
\end{center}
\caption{Parallel plates.  The dashed line indicates the `gap' between
the plates where the non-local field theory lives.  There is also a
periodic spatial dimension of size $L_z \rightarrow \infty$ pointing out
of the page and a periodic Euclidean time dimension of size $\beta$.
\label{fig:parallel}}
\end{figure}

In this section we study the low temperature behavior of the free energy
for parallel plates.  The geometry is shown in Fig.~\ref{fig:parallel}.

\noindent
As before, there are three contributions to the free energy.

\noindent
{\em Bulk contribution} \\
The bulk contribution to the free energy (\ref{bulk}) has two
components.  In the region above the middle plate we have an ideal gas
in infinite volume, with a free energy given in (\ref{4dhigh}).  In
the region below the middle plate we have an ideal gas at low
temperature ($bT \ll 1$), with a thermal free energy given in (\ref{4dlow}) that is exponentially
suppressed.  Overall we have
\be
\label{ParallelBulk}
F^{\rm bulk}_{||,T} = - {\zeta(4) \over \pi^2} \, V_{\rm top} T^4 + {\zeta(3) \over 8 \pi} \, A_{\rm top} T^3
- {\zeta(2) \over 16 \pi} \, P_{\rm top} T^2
\ee
To make this well defined we are actually working with the geometry
shown in Fig.~\ref{fig:HalfGeometry} in the limit $a,b_1 \rightarrow \infty$
with $b = b_2$ fixed.  The quantities $V_{\rm top}$, $A_{\rm top}$,
$P_{\rm top}$ refer to the volume, surface area, and ``perimeter'' (length of the corners)
of the region above the middle plate.  For instance $P_{\rm top} = 4 L_z$.

\medskip
\noindent
{\em Direct contribution} \\
The direct contribution to the free energy is 
\bea
\label{Z2dDirect}
F^{\rm direct}
& = & {1 \over {2 \beta}} {\rm Tr} \log \big({\cal O}_{\rm top}^{\rm direct} + {\cal O}_{\rm bottom}^{\rm direct}\big) \\
\nonumber
& = & {1 \over {2 \beta}} {\rm Tr} \log \left[\sqrt{(n \pi / a)^2 + \mu^2}
\left(1 + \coth \left(b \sqrt{(n \pi / a)^2 + \mu^2}\right)\right)\right]
\eea
Here $b = b_2$ is the distance between the plates.  We have set $b_1 =
\infty$ but kept $a$ as an infrared regulator.

The direct contribution breaks up into two pieces.  The first piece is
\be
F^{(1)\,\rm direct}_{||} = {1 \over 2\beta} \sum_{n,k,l} \log \left(2  \sqrt {\big({{n \pi} \over {a}}\big)^2
+\big({{2 k \pi} \over L_z}\big)^2 + \big({{2 l \pi} \over \beta}\big)^2 }\right)
\ee
This is half the free energy of an ideal gas in $2+1$ dimensions, in a box with a Dirichlet direction of
size $a$ and a periodic direction of size $L_z$.  This is evaluated in appendix \ref{sect:thermo}, equation (\ref{3dhigh}).
We find
\be
\label{term1}
F^{(1)\,\rm direct}_{||,T} = - {\zeta(3) \over 4 \pi} \, L_z a T^3 + {\zeta(2) \over 4 \pi} \, L_z T^2
\ee
The second piece of the free energy is
\be
\label{E2}
F^{(2)\,\rm direct}_{||}  =  - {1 \over {2 \beta}}  \log \left(1 - e^{-2b\sqrt {( {{n \pi} \over {a}})^2+({{2k \pi} \over L_z})^2 +({{2 l  \pi} \over \beta})^2 }}\right)
\ee
This is studied in appendix \ref{sect:direct}, equation (\ref{term2a}).  At low temperatures, $bT \ll 1$, we find
\be
\label{term2}
F^{(2)\,\rm direct}_{||,T} = - {\zeta(2) \over 4 \pi} \, L_zT^2 + {\zeta(3) \over 4\pi} L_z(a+b)T^3 - {\zeta(4) \over \pi^2} L_zabT^4
\ee
Combining (\ref{term1}) and (\ref{term2}) there are some cancellations,
leaving
\be
\label{ParallelDirect}
F^{\rm direct}_{||,T} = - {\zeta(4) \over \pi^2} \, L_zabT^4 + {\zeta(3) \over 4 \pi} L_zbT^3
\ee

\medskip
\noindent
{\em Diffractive contribution} \\
Finally we turn to the diffractive contribution (\ref{diffractive}).
Combining the top and bottom contributions we have the direct matrix elements
\be
{\cal O}^{\rm direct}_{mn} = 2 \sqrt{(m \pi / a)^2 + \mu^2} \left(1 - e^{-2 b \sqrt{(m \pi / a)^2 + \mu^2}}\right)^{-1} \delta_{mn}
\ee
The top diffractive matrix element is, sending $a,b \rightarrow \infty$ in (\ref{OddDiffractive}),
\be
{\cal O}^{\rm top,\,diffractive}_{mn} = - {2 \mu^2 \over \pi a^3} \int_1^\infty dy \, \sqrt{y^2 - 1} \,
{m \pi \over (m \pi / a)^2 + \mu^2 y^2} \, {n \pi \over (n \pi / a)^2 + \mu^2 y^2}
\ee
where the sum became an integral over $y = \sqrt{(j \pi / \mu b)^2 +
1}$.  The bottom diffractive matrix element is
\bea
\nonumber
{\cal O}_{mn}^{\rm bottom,\,diffractive} &=& - 2 a b \sum_{j = 1}^\infty
\left(1 - \exp\Big(-2 a \sqrt{(j \pi / b)^2 + \mu^2}\Big)\right)
{j^2 \pi^2 \over \sqrt{(j \pi / b)^2 + \mu^2}} \\
&& \qquad {m \pi \over (m \pi b)^2 + (j \pi a)^2 + (\mu a b)^2} \,\,
                        {n \pi \over (n \pi b)^2 + (j \pi a)^2 + (\mu a b)^2}
\eea

The bottom diffractive contribution can be obtained from previous results.  Note that ${1 \over 2} {\rm Tr} \, {\cal O}_{\rm direct}^{-1}
{\cal O}_{\rm bottom,\,diffractive}$ is symmetric under exchange of $a$ and $b$.  As $a \rightarrow \infty$ it can be analyzed along the lines
of the first diffractive contribution for perpendicular plates.  In fact it gives exactly half of the
perpendicular plate result (\ref{PerpZ2d}) with the replacement $a \rightarrow b$.  So from (\ref{F4dPerp})
it makes a contribution $-{8 \pi \zeta(3) \over 945} L_z b^4 T^6$ to the free energy in four dimensions.
This will turn out to be a subleading contribution at low temperatures.

The leading diffractive contribution to the 2d partition function comes from the top matrix elements.
\bea
\nonumber
-\log Z_{\rm 2d}^{\rm diffractive} & = & {1 \over 2} {\rm Tr} \, {\cal O}_{\rm direct}^{-1} {\cal O}_{\rm top,\,diffractive} \\
\label{DiffractiveInt}
& = & - {1 \over 4\pi^2} \int_0^1 dz \, \left(1 - e^{-2 \mu b / z}\right) g(z)
\eea
To obtain this we did the integral over $y$, the trace became
an integral over $z = \mu/ \sqrt{(n \pi / a)^2 + \mu^2}$, and we introduced the function
\be
g(z) = {1 \over z \sqrt{1 - z^2}} - {z \over 1 - z^2} \, {\rm cosh}^{-1}(1/z)
\ee
It is convenient to break the integral (\ref{DiffractiveInt}) into two pieces.  The first piece is
\be
- \log Z_{\rm 2d}^{\rm divergent} = - {1 \over 4\pi^2} \int_0^1 dz \, g(z)
\ee
This is log divergent since $g(z) \sim 1/z$ at small $z$.  We can regulate
the divergence by introducing a momentum cutoff $\Lambda$ (a cutoff on the value
of $n \pi / a$).  This corresponds to a lower limit of integration at $z = \mu/\sqrt{\Lambda^2
+ \mu^2}$.  The regulated contribution to the partition function is then
\bea
\nonumber
- \log Z_{\rm 2d}^{\rm divergent} & = & - {1 \over 4\pi^2} \int_{\mu/\sqrt{\Lambda^2
+ \mu^2}}^1 dz \, g(z) \\
\label{logZ2ddiv}
& = & - {1 \over 4\pi^2} \log {2 \Lambda \over \mu} + {1 \over 32} + {\cal O}\big(1/\Lambda^2\big)
\eea
The second piece of (\ref{DiffractiveInt}) is
\be
- \log Z_{\rm 2d}^{\rm finite} = {1 \over 4\pi^2} \int_0^1 dz \, e^{-2 \mu b / z} g(z)
\ee
To determine the non-analytic behavior as $\mu b \rightarrow
0$ we introduce a separation scale $z_0$ and break the
integral over $z$ up into ultraviolet ($0 < z < z_0$) and infrared
($z_0 < z < 1$) regions.  The choice of separation scale is a bit
subtle since it has to scale with $\mu b$ as $\mu b \rightarrow 0$.
The correct prescription is to set $z_0 = c \sqrt{\mu b}$ where $c$
is an arbitrary constant.  The ultraviolet contribution is then
\bea
\label{logZ2dUV}
- \log Z_{\rm 2d}^{\rm UV} & = & {1 \over 4\pi^2} \int_0^{z_0} dz \, e^{-2 \mu b / z} g(z) \\
\nonumber
& = & {1 \over 4\pi^2} \int_0^{z_0} dz \, e^{-2 \mu b / z}
\left({1 \over z} + z ({1 \over 2}+\log {z \over 2} ) + z^3 ( {5 \over 8} + \log {z \over 2} ) + {\cal O}\big(z^5\big)\right)
\eea
Integrating term-by-term gives the ultraviolet contribution as an expansion in powers of $\sqrt{\mu b}$.
Likewise the infrared contribution is
\bea
\label{logZ2dIR}
- \log Z_{\rm 2d}^{\rm IR} & = & {1 \over 4\pi^2} \int_{z_0}^1 dz \, e^{-2 \mu b / z} g(z) \\
\nonumber
& = & {1 \over 4\pi^2} \int_{z_0}^1 dz \, \left(1 - {2 \mu b \over z} + {2 (\mu b)^2 \over z^2}
- {4 (\mu b)^3 \over 3 z^3} + {2 (\mu b)^4 \over 3 z^4} + {\cal O}\big(1/z^5\big)\right) g(z)
\eea
Again integrating term-by-term gives an expansion in powers of $\sqrt{\mu b}$.
Putting (\ref{logZ2dUV}) and (\ref{logZ2dIR}) together we find
\bea
\nonumber
- \log Z_{\rm 2d}^{\rm finite} & = & - {1 \over 4 \pi^2} \log (\mu b) + {1 \over 8} \mu b
- {1 \over 4 \pi^2} (\mu b)^2 \left[\log^2 (\mu b) + 2 (\gamma - 1) \log(\mu b)\right]  \\
\label{logZ2dfinite}
& & \qquad + \, {1 \over 12} (\mu b)^3 + \cdots
\eea
In this expression $\cdots$ denotes higher-order non-analytic terms as well as odd analytic terms of order $(\mu b)^5$ and higher.  Terms analytic in $\mu^2$ have been neglected since they give exponentially small thermal corrections. Note that the dependence on
$c$ cancels between the UV and IR contributions, and the final expression (\ref{logZ2dfinite}) does not depend on $c$.  Putting
(\ref{logZ2ddiv}) and (\ref{logZ2dfinite}) together we have\footnote{Again we have dropped terms analytic in $\mu^2$.  This includes the cutoff
dependence which only appears in the combination $\log (\Lambda b)$.}
\be
-\log Z_{\rm 2d}^{\rm diffractive} = {1 \over 8} \mu b
- {1 \over 4 \pi^2} (\mu b)^2 \left[\log^2 (\mu b) + 2 (\gamma - 1) \log(\mu b)\right] + {1 \over 12} (\mu b)^3 + \cdots 
\ee
From (\ref{Z4dthermal}) the low temperature behavior of the 4d
free energy is then
\bea
\label{ParallelDiffractive}
F^{\rm diffractive}_{||,T} & = & { L_z \over \pi b^2} \sum_{l = 1}^\infty \int_0^\infty d(\mu b)
J_0(\beta l \mu) \left(- \mu b \log Z_{\rm 2d}^{\rm diffractive}\right) \nonumber\\
& = & {L_z \over {b^2}} \left[ - {\zeta(3) \over 8 \pi} (bT)^3 - {2  (b T)^4 \over \pi^3}
\sum_{l=1}^\infty {1 \over l^4} \log (2bT/l) + {{3 \zeta(5)} \over {4 \pi}} (bT)^5 + \cdots \right] \nonumber \\
&= & {L_z \over {b^2}} \left[ - {\zeta(3) \over 8 \pi} (bT)^3 - {2 \zeta(4) \over \pi^3} \, (bT)^4 \left(\log(2bT) + {\zeta'(4) \over \zeta(4)}\right)+ {{3 \zeta(5)} \over {4 \pi}} (bT)^5 + \cdots \right] \nonumber \\
\eea
The diffractive contribution to the free energy is purely an edge effect.

\begin{figure}[h]
\begin{center}
\includegraphics[width = 7.8cm]{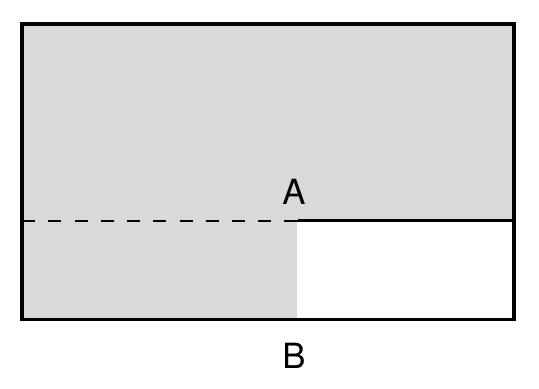}
\end{center}
\caption{At low temperature the free energy for parallel plates (\ref{ParallelCombined})
comes in part from an ideal gas filling the shaded region.
\label{fig:Veff}}
\end{figure}

\medskip
\noindent
{\em Summary} \\
Combining (\ref{ParallelBulk}), (\ref{ParallelDirect}) and (\ref{ParallelDiffractive}) we have
\bea
F_{||,T} &=& - {\zeta(4) \over \pi^2} \, V_{\rm eff} T^4 + {\zeta(3) \over 8 \pi} A_{\rm eff} T^3
- {\zeta(2) \over 16 \pi} P_{\rm eff} T^2 \nonumber \\
\label{ParallelCombined}
&&- {2 \zeta(4) \over \pi^3} \,
L_zb^2T^4 \left(\log(2bT) + {\zeta'(4) \over \zeta(4)}\right)+{{3 \zeta(5)} \over {4 \pi}} L_z b^3T^5 + \cdots 
\eea
The first three terms have a simple geometrical interpretation, as the
free energy of an ideal gas filling the shaded region in
Fig.~\ref{fig:Veff}.  Here $V_{\rm eff} = V_{\rm top} + L_zab$ is the volume
of the shaded region, while $A_{\rm eff} = A_{\rm top} + L_zb$ is its
effective surface area and $P_{\rm eff} = 4 L_z$ is its effective
perimeter.  There are some important cancellations that go into this
result.  In particular, due to a partial cancellation between
(\ref{ParallelDirect}) and (\ref{ParallelDiffractive}), $A_{\rm eff}$
only counts the surface area of the shaded region associated with solid lines in
Fig.~\ref{fig:Veff}.  Also the two extra corners of the shaded region
(denoted $A$ and $B$ in the figure) do not contribute to $P_{\rm eff}$.
The final term in the free energy is a purely diffractive effect and does
not have a simple geometric interpretation.

A nice way to interpret this result is to isolate the thermal Casimir energy associated with the
gap.  Proceeding as in section \ref{sect:perp} we first subtract the free energy of a ``big box''
without any middle plate, given by
\be
F_{\rm box} = -{\zeta(4) \over \pi^2} V_{\rm box} T^4 + {\zeta(3) \over 8 \pi} A_{\rm box} T^3
- {\zeta(2) \over 16\pi} P_{\rm box} T^2
\ee
Next we subtract the thermal self-energy of the middle plate itself, as well as the thermal self-energy associated
with the ``$\dashv\,$'' shaped junction on the right side of Fig.~\ref{fig:Veff}.  These are give by
\be
F_{\rm self}
= {\zeta(3) \over 4 \pi} A_{\rm plate} T^3 - {\zeta(2) \over 8 \pi} L_z T^2
\ee
The bulk contribution to the free energy associated with the gap in the middle plate is then
\bea
\label{ParallelSummary}
F_{||,T} - F_{\rm box} - F_{\rm self} & = & {\zeta(4) \over \pi^2} \, V_{\rm ex} T^4 - {\zeta(3) \over 8 \pi}
A_{\rm ex} T^3 + {\zeta(2) \over 16 \pi} P_{\rm ex} T^2 \\
\nonumber
& & \hspace{-3mm} - {2 \zeta(4) \over \pi^3} \,
L_zb^2T^4 \left(\log(2bT) + {\zeta'(4) \over \zeta(4)}\right) + {{3 \zeta(5)} \over {4 \pi}} L_z b^3T^5 + \cdots 
\eea
Here $V_{\rm ex}$ is the excluded volume (the volume of the region between the two plates, shown in white in
Fig.~\ref{fig:Veff}).  Likewise $A_{\rm ex} = 2 A_{\rm plate} + bL_z$ is the excluded area (the surface area
of the region in white, counting just the boundaries with solid lines), and $P_{\rm ex} = 2 L_z$ is the excluded
perimeter.  These geometrical terms have a simple interpretation, that at low temperatures
thermal excitations cannot propagate in the region between the plates.

The leading diffractive contribution to the thermal free energy associated with the edge is
\be
\label{edge}
F^{\rm edge}_{||,T} = - {2 \zeta(4) \over \pi^3} \,
(bT)^4 \left(\log(2bT) + {\zeta'(4) \over \zeta(4)}\right) {L_z \over{b^2}}
\ee
This contribution to the thermal free energy was studied by Gies and Weber in  \cite{GiesWeber}
using the world-line formalism. They observed that their numerical data was well fit, in the low temperature limit, by a power-law temperature dependence with a non-integer exponent $\sim T^{3.74}$. A numerical fit of our analytic result (\ref{edge}) in terms of a power-law dependence, for low temperatures, agrees well with the data in \cite{GiesWeber} and produces a similar exponent. However it is clear from our analysis that the
non-integer power law found in \cite{GiesWeber} is actually due to a logarithmic temperature
dependence of the form $T^4 \log T$.

\section{Conclusions\label{sect:conclusions}}

To summarize, we find that up to first order in diffractive effects, the thermal free energy at low temperature is

\medskip
\noindent
{\em Perpendicular plates} \\
\be
F_{\perp, T} = - {\zeta(2) \over 8 \pi} L_z T^2 + {\zeta(3) \over 4\pi} L_z a T^3 - {16 \pi \zeta(3) \over 945} L_z a^4 T^6 + {\cal O}(T^8)
\ee
This was given in (\ref{PerpSummary}).  

\medskip
\noindent
{\em Slit geometry} \\
\be
F_{\rm slit, T} =  {\zeta(3) \over 4\pi} L_z w T^3 - {7  \zeta(3) \over {90 \pi}} L_z w^2 T^4 + {\cal O}(T^6)
\ee
as given in (\ref{SlitSummary}).

\medskip
\noindent
{\em Parallel plates} \\
For parallel plates we find (\ref{ParallelSummary}), which can be decomposed into an excluded volume contribution
\be
F_{||,T}^{\rm ex} = {\zeta(4) \over \pi^2} \, V_{\rm ex} T^4 - {\zeta(3) \over 8 \pi} A_{\rm ex} T^3 + {\zeta(2) \over 16 \pi} P_{\rm ex} T^2
\ee
and a diffractive edge contribution
\be
\label{EdgeConclusion}
F^{\rm edge}_{||,T} = - {2 \zeta(4) \over \pi^3} \, (bT)^4 \left(\log(2bT) + {\zeta'(4) \over \zeta(4)}\right) {L_z \over{b^2}}+ {{3 \zeta(5)} \over {4 \pi}} L_z b^3T^5 + \cdots 
\ee

These results are consistent with the world-line numerical analysis in \cite{KlingGies,GiesWeber} and they further capture subleading temperature dependence arising from diffractive effects. The result (\ref{EdgeConclusion}) provides an analytic understanding of the fractional
power law observed in \cite{GiesWeber}.  From a mathematical point of view we find it interesting that these non-trivial
power laws are encoded in the non-local differential operators (\ref{operators}).

Our method is rather general and can be applied to many contexts in
field theory where geometrical and thermal effects, and in particular
the interplay between them, are important. For instance they could be
used to study thermal corrections to the interaction between holes in
a plate \cite{Kabat:2010yy}.  It is also straightforward to extend our
results to higher dimensions.  Another analytical approach to studying
Casimir energies in geometries with edges and apertures is the
multiple scattering method developed in \cite{MIT}. It would be
interesting to understand the relation between the expansion scheme
developed here and the methods used in \cite{MIT}, as well as the
convergence properties of these expansions at any temperature.

\bigskip
\goodbreak
\centerline{\bf Acknowledgements}
\noindent
We are grateful to V.P.\ Nair for valuable discussions, and we thank Noah Graham, Robert Jaffe and Mohammad Maghrebi for hospitality and
stimulating comments.  This work was supported by U.S.\ National Science Foundation grants PHY-0855582 and PHY-0758008 and
by PSC-CUNY grants.

\appendix
\section{Ideal gas thermodynamics}\label{sect:thermo}

The partition function for an ideal gas in a rectangular box of size $L_x \times b \times L_z$, with Dirichlet boundary conditions in the $L_x$ and $b$ directions and periodic boundary conditions around $L_z$ and $\beta$, is 
\bea
\label{A1}
- \log Z_{\rm 4d} & = & {1 \over 2} {\rm Tr} \log \big(-\Box\big) \nonumber \\
& = & {1 \over 2} \sum  \log \left[ \left({{n \pi} \over L_x}\right)^2 + \left({m \pi \over b}\right)^2+\left({2 k \pi \over L_z}\right)^2 +\left({2 l \pi \over \beta}\right)^2 \right]
\eea
where $n,m=1,2,\cdots$ and $k,l \in {\mathbb Z}$.  As discussed in \cite{Kabat:2010nm} appendix B, the renormalized
partition function is, in the limit $L_x,\,L_z \rightarrow \infty$,
\be
\label{A2}
- \log Z_{\rm 4d} = - {1 \over 2} \int_0^\infty {ds \over s} \, \left({L_x \over \sqrt{4 \pi s}} - {1 \over 2}\right) {L_z \over \sqrt{4 \pi s}} \, \left[
K_P(s,\beta) K_D(s,b) - {\beta \over \sqrt{4 \pi s}} \left({b \over \sqrt{4 \pi s}} - {1 \over 2}\right) \right]
\ee
where the heat kernels associated with periodic (P) and Dirichlet (D) directions are
\bea
\label{KP}
K_P(s,\beta) & = & {\beta \over \sqrt{4 \pi s}} + {\beta \over \sqrt{\pi s}} \sum_{n = 1}^\infty e^{-\beta^2 n^2 / 4s} \\
\label{KD}
K_D(s,b) & = & {b \over \sqrt{4 \pi s}} - {1 \over 2} + {b \over \sqrt{\pi s}} \sum_{n = 1}^\infty e^{-b^2 n^2 / s}
\eea
The expansions (\ref{KP}), (\ref{KD}) are useful when $\beta$ or $b$ are large.  For small $\beta$ or $b$ we use the Poisson-resummed forms
\bea
\label{KP2}
K_P(s,\beta) & = & 1 + 2 \sum_{n = 1}^\infty e^{-s 4 \pi^2 n^2 / \beta^2} \\
\label{KD2}
K_D(s,b) & = & \sum_{n = 1}^\infty e^{-s n^2 \pi^2 / b^2}
\eea

To study the behavior at low temperature ($\beta \gg b$) we rewrite (\ref{A2}) as
\bea
\nonumber
- \log Z_{\rm 4d} &=& - {1 \over 2} \int_0^\infty {ds \over s} \, \left({L_x \over \sqrt{4 \pi s}} - {1 \over 2}\right) {L_z \over \sqrt{4 \pi s}} \,
\left[\left(K_P(s,\beta) - {\beta \over \sqrt{4 \pi s}}\right) K_D(s,b) \right. \\
\label{A3}
& & \hspace{3cm} + \left.{\beta \over \sqrt{4 \pi s}} \left(K_D(s,b) - {b \over \sqrt{4 \pi s}} + {1 \over 2}\right)\right]
\eea
From (\ref{KP}), (\ref{KD2}) the first line is exponentially suppressed at low temperatures, while the second line can be evaluated analytically.  After
integrating over $s$ we find
\bea
\nonumber
- \log Z_{\rm 4d} &=& - {\zeta(4) \beta L_x L_z \over 16 \pi^2 b^3} + {\zeta(3) \beta L_z \over 32 \pi b^2}
- {L_x L_z \over \sqrt{2 \beta b^3}} \sum_{m,n = 1}^\infty \left({m \over n}\right)^{3/2} K_{3/2}\left(mn\pi\beta/b\right) \\
\label{4dlow}
&& + {L_z \over 2 b} \sum_{m,n = 1}^\infty {m \over n} K_1\left(mn\pi\beta/b\right)
\eea
The first two terms determine the Casimir energy at zero temperature associated with this geometry,
\be
E^{\rm Casimir}_{T = 0} = - {\zeta(4) L_x L_z \over 16 \pi^2 b^3} + {\zeta(3) L_z \over 32 \pi b^2}
\ee
while the remaining terms give exponentially small thermal corrections.

To study the behavior at high temperatures ($\beta \ll b$) we rewrite (\ref{A2}) as
\bea
\nonumber
- \log Z_{\rm 4d} &=& - {1 \over 2} \int_0^\infty {ds \over s} \, \left({L_x \over \sqrt{4 \pi s}} - {1 \over 2}\right) {L_z \over \sqrt{4 \pi s}} \,
\left[K_P(s,\beta) \left(K_D(s,b) - {b \over \sqrt{4 \pi s}} + {1 \over 2} \right) \right. \\
\label{A4}
& & \hspace{3cm} + \left.\left(K_P(s,\beta) - {\beta \over \sqrt{4 \pi s}}\right)\left({b \over \sqrt{4 \pi s}} - {1 \over 2}\right)\right]
\eea
We use (\ref{KP2}), (\ref{KD}) in the first line, while the second line can be evaluated analytically.  Thus
\bea
\label{4dhigh}
- \log Z_{\rm 4d} & = & - {\zeta(4) \over \pi^2} V T^3 + {\zeta(3) \over 8 \pi} A T^2 - {\zeta(2) \over 16 \pi} PT - {\zeta(3) L_x L_z \over 16 \pi b^2} + {\zeta(2) L_z \over 8 \pi b}  \\[2pt]
\nonumber
& & - L_x L_z \sqrt{2 \over b \beta^3} \sum_{m,n = 1}^\infty \left({m \over n}\right)^{3/2} K_{3/2}\left(mn4\pi b/\beta\right)
+ {L_z \over \beta} \sum_{m,n = 1}^\infty {m \over n} K_1\left(mn4\pi b/\beta\right)
\eea
Here $V = L_x b L_z$ is the volume of the box, $A = 2(L_x + b) L_z$ is the surface area, and $P = 4 L_z$ is the ``perimeter''
(the length of the corners).  The terms which are independent of $T$ come from $K_P(s,\beta) = 1 + \cdots$ in the first line; they give the Casimir energy associated
with this geometry after dimensional reduction along the Euclidean time direction.  The volume term in (\ref{4dhigh}) gives the usual extensive
free energy of an ideal gas; note that only Dirichlet boundaries count towards the surface area.

One can perform a similar analysis in 2+1 dimensions.  For a gas in a box of size $b \times L_z$, with Dirichlet boundary conditions in $b$ and periodic
boundary conditions around $L_z$ and $\beta$, the starting point is, for $L_z \rightarrow \infty$,
\be
\label{A5}
- \log Z_{\rm 3d} = - {1 \over 2} \int_0^\infty {ds \over s} \, {L_z \over \sqrt{4 \pi s}} \, \left[
K_P(s,\beta) K_D(s,b) - {\beta \over \sqrt{4 \pi s}} \left({b \over \sqrt{4 \pi s}} - {1 \over 2}\right) \right]
\ee
Proceeding as before, at low temperatures we have
\be
\label{3dlow}
- \log Z_{\rm 3d} = - {\zeta(3) \beta L_z \over 16 \pi b^2} - {L_z \over b} \sum_{m,n = 1}^\infty {m \over n} K_1(m n \pi \beta / b)
\ee
The first term gives the Casimir energy at zero temperature in 2+1 dimensions, while the remaining terms are exponentially
small thermal corrections.  At high temperatures the steps leading to (\ref{4dhigh}) give
\be
\label{3dhigh}
- \log Z_{\rm 3d} = - {\zeta(3) b L_z \over 2 \pi \beta^2} + {\pi L_z \over 12 \beta} - {\pi L_z \over 24 b} - {2 L_z \over \beta} \sum_{m,n = 1}^\infty {m \over n} K_1(m n 4 \pi b / \beta)
\ee

\section{Direct free energy for parallel plates}\label{sect:direct}

In this appendix we compute the second piece of the direct free energy for parallel plates (\ref{E2}).
The 2d partition function is
\beas
-\log Z_{\rm 2d}^{(2)} & = & - {1 \over 2} \sum_{n=1}^\infty \log \left(1 - e^{-2b\sqrt{(n \pi / a)^2 + \mu^2}}\right) \\
& = & - {a \over 2\pi} \int_0^\infty dk \, \log \left(1 - e^{-2b\sqrt{k^2 + \mu^2}}\right) + {1 \over 4} \log
\left(1 - e^{-2b\mu}\right) + {\cal O}(1/a)
\eeas
where we used the Euler-Maclaurin summation formula to obtain the
behavior for large $a$.  Letting $x = b \sqrt{k^2 + \mu^2}$ and
integrating by parts this is
\be
\label{SecondTerm}
{a \over \pi b} \int_{\mu b}^\infty dx \, \sqrt{x^2 - \mu^2 b^2} \, \left(e^{2 x} - 1\right)^{-1}
+ {1 \over 4} \log \left(1 - e^{-2b\mu}\right) + {\cal O}(1/a)
\ee
The non-analytic behavior of the integral as $\mu b \rightarrow 0$ can
be obtained by the method explained below (\ref{PerpZ2d}).  Keeping only
terms which are non-analytic as functions of $\mu^2$, we find that
\be
- \log Z_{\rm 2d}^{(2)} = {1 \over 4} \log \mu b - {1 \over 4} \mu (a + b) - {1 \over 4 \pi} a b \mu^2 \log \mu b
+ {\cal O}\big((\mu b)^4\big)
\ee
Substituting this in (\ref{Z4dthermal}), the four dimensional free energy is
\bea
\nonumber
 F_{\rm 4d}^{(2)} & = & {L_z \over \pi} \sum_{l = 1}^\infty \int_0^\infty \mu d\mu \, J_0(\beta \mu l)
\left({1 \over 4} \log \mu b - {1 \over 4} \mu (a + b) - {1 \over 4 \pi} a b \mu^2 \log \mu b\right) \\
\label{term2a}
& = & - {\zeta(2) \over 4 \pi} \, L_zT^2 + {\zeta(3) \over 4\pi} L_z(a+b)T^3 - {\zeta(4) \over \pi^2} L_zabT^4
\eea

Another approach to evaluating (\ref{E2}) is to begin from the partition function for a Bose gas in a box of size $a \times L_z \times 2b$,
with Dirichlet boundary conditions in $a$ and periodic boundary conditions around $L_z$ and $2b$.  With $a,L_z \rightarrow \infty$ this is
\be
\label{ZBose}
- \log Z_{\rm Bose} = - {1 \over 2} \int_0^\infty {ds \over s} \, \left({a \over \sqrt{4 \pi s}} - {1 \over 2}\right) {L_z \over \sqrt{4 \pi s}} \, \left[
K_P(s,\beta) K_P(s,2b) - {\beta \over \sqrt{4 \pi s}} {2b \over \sqrt{4 \pi s}} \right]
\ee
This partition function is manifestly symmetric under exchange $\beta \leftrightarrow 2b$.
Evaluating it at large $\beta$ as in appendix \ref{sect:thermo}, we find
\bea
\label{ZBoseLow}
- \log Z_{\rm Bose} &=& - {\zeta(4) a L_z \beta \over 8 \pi^2 b^3} + {\zeta(3) L_z \beta \over 16 \pi b^2} - {\zeta(3) a L_z \over 2 \pi \beta^2} + {\zeta(2) L_z \over 2 \pi \beta} \\
\nonumber
&& + (\hbox{\rm exponentially small thermal corrections})
\eea
while evaluating it at small $\beta$ gives
\bea
\label{ZBoseHigh}
- \log Z_{\rm Bose} &=& - {2\zeta(4) a L_z b \over \pi^2 \beta^3} + {\zeta(3) L_z b \over 2 \pi \beta^2} - {\zeta(3) a L_z \over 8 \pi b^2} + {\zeta(2) L_z \over 4 \pi b} \\
\nonumber
&& + (\hbox{\rm exponentially small finite size corrections})
\eea
Regarding $2b$ as the Euclidean time direction and working in a Hamiltonian picture we have
\be
- \log Z_{\rm Bose} = {\rm Tr} \log \left[2 \sinh\big(b\sqrt {(n \pi/a)^2+(2k \pi/L_z)^2 +(2 l  \pi/\beta)^2}\big)\right]
\ee
After multiplying by an overall factor of $-1/2$, this can be identified with the contribution (\ref{E2}) to the direct free energy for parallel plates,
except that in (\ref{E2}) the zero point energy has been suppressed.  That is, we can identify
\be
\label{NoCasimir}
\beta F^{(2)}_{\rm 4d} = - {1 \over 2} \left(- \log Z_{\rm Bose} - 2 b E_{\rm Casimir}\right)
\ee
where the Casimir energy at zero ``temperature'' (meaning $2b \rightarrow \infty$) for this geometry is, from (\ref{ZBoseHigh}),
\be
\label{ECasimir}
E_{\rm Casimir} = - {\zeta(4) aL_z \over \pi^2\beta^3} + {\zeta(3) L_z \over 4 \pi \beta^2}
\ee
Using (\ref{ZBoseLow}), we find that (\ref{NoCasimir}) reproduces the temperature dependence seen in (\ref{term2a}),
and in fact shows that corrections to (\ref{term2a}) are exponentially small.


\providecommand{\href}[2]{#2}\begingroup\raggedright\endgroup

\end{document}